# Nonlinear optics driven magnetism reorientation in semiconductors

Qianqian Xue, Yan Sun, Jian Zhou[*]

*Center for Alloy Innovation and Design, State Key Laboratory for Mechanical Behavior of Materials, Xi'an Jiaotong University, Xi'an 710049, China*

Abstract

Based on nonlinear optics, we propose that light irradiation could induce a steady state magnetization variation. We develop a band formulism to elucidate its general microscopic mechanisms, which are rooted by the quantum geometric structure and topological nature of electronic Bloch wavefunctions. Their existence are determined by the light polarization and specific material symmetry, based on the magnetic group theory. In general, for a magnetic system, both circularly and linearly polarized light could exert an effective magnetic field and a magnetic velocity (variation rate over time, serving as an effective torque) effect, to reorient the magnetization orientation. They are contributed by spin and orbital angular momenta simultaneously. Aided by group theory and first-principles calculations, we illustrate this theory using a showcase example of monolayer $NiCl_2$, showing that light irradiation effectively generates an out-of-plane effective magnetic torque, which lifts its in-plane easy magnetization. According to magnetic dynamic simulations, the in-plane magnetization could be switched to the out-of-plane direction in a few nanoseconds under a modest light intensity, demonstrating its ultrafast nature desirable for quantum manipulation.

[*] Corresponding author: jianzhou@xjtu.edu.cn

**Introduction**

Light-matter interaction is essential for many ultrafast and exotic physical processes and applications, owing to its non-contacting and non-invasive nature. While light induced phase transformation [1-4] and current generation [5-8] mainly involve mechanical, thermal, and electrical responses, generation and manipulation magnetic orders via light has received increasing attentions over the past a few decades [9-11]. Magnetic materials play one of the most fundamental roles in the modern science and technology, including magnetic memory and storage [12,13], information process [14-16], ultrafast detection [17,18] and sensor applications [19-22], etc. Among them, manipulating the magnetization is crucial for various applications [23-25]. Modern technology mainly relies on current induced magnetization switching, according to the inverse Rashba-Edelstein effect [26,27] and inverse spin Hall effect [28,29]. It remains an open question whether light is capable of switching the magnetization of intrinsic magnets, though light-induced static magnetization has been intensively studied such as the inverse Faraday [30,31] and inverse Cotton-Mouton effects [32,33].

In this work, we apply the nonlinear optics approach to demonstrate the feasibility of light-harnessed magnetization. In order to achieve static (zero frequency) magnetization switch under optical field (angular frequency $\omega$), the lowest order response is quadratic [34,35]. We develop a general band perturbation theory for magnetic semiconductor materials to show how light can effectively switch the intrinsic magnetization. This switching process arises from interband transitions and is parity-even, in stark contrast to the symmetry requirement of the Rashba-Edelstein effect or magnetoelectric coupling. We further enumerate this process into four different categories, under linearly polarized light (LPL) and circularly polarized light (CPL), and present a group theory analysis on their symmetry conditions. They arise from geometric nature of Bloch wavefunctions at the valence and conduction bands, featuring their topological nature. In order to illustrate this process, we showcase our theory in a typical 2D ferromagnetic material, monolayer $NiCl_2$ which prefers an in-plane magnetic moment [36]. Both CPL and LPL could generate an out-of-plane effective magnetic field ($\boldsymbol{B}_{\text{eff}}$) and magnetic velocity (serving as an effective magnetic torque $\boldsymbol{T}$). They contain contributions from spin ($\boldsymbol{S}$) angular momentum (SAM) and orbital ($\boldsymbol{L}$) angular momentum (OAM). The latter is reminiscent of the orbital Hall effect which had long been overlooked largely, but recently attracts tremendous revived attention [37,38]. By performing magnetic dynamics simulations [39], we illustrate that such photoinduced magnetization reorientation can occur on the order of 1 ns when the magnet is under moderate light shining, thus subject to an ultrafast quantum manipulation process.

**Results**

**A. Band theory of nonlinear optical control over magnetization**

It is well-known that under static electric-field, the Rashba-Edelstein effect or magnetoelectric coupling could induce variation of SAM, whose magnitude can be derived according to the linear response theory,

$$\delta S^a = \chi^a_b E_b. \tag{1}$$

Here, the coefficient $\chi^a_b$ measures how the $b$-component of the electric field $\boldsymbol{E}$ changes the SAM component $\delta S^a$. According to the band theory [26,40-42], it mainly arises from the intraband transition at the Fermi surface, while the Fermi sea contribution is usually marginal. The time-reversal ($\mathcal{T}$) even components is proportional to the carrier lifetime ($\chi^a_b \sim \tau$), similar as the Drude model [43]. Under spatial inversion ($\mathcal{P}$), one has $\mathcal{P} E_b = -E_b$ and $\mathcal{P} S^a = S^a$, indicating that $\chi^a_b$ is $\mathcal{P}$-odd. Hence, the Rashba-Edelstein effect poses $\mathcal{P}$-broken requirements for its material platforms, and typically also requires metallicity to mediate intraband transitions.

We propose that such constraints can be lifted when one adopts light illumination as an alternative route to trigger SAM variation. Under the alternating electric field of light $\boldsymbol{\mathcal{E}}(\omega) = \boldsymbol{E} e^{i\omega t}$, the lowest order response to generate a static magnetization switch is quadratic, similar as the static current generation in the bulk photovoltaic effect [34,35,44,45]. The response function can be written as

$$\delta S^a = \sum_{\Omega=\pm\omega} \chi^a_{bc} \mathcal{E}_b(\Omega) \mathcal{E}_c(-\Omega), \tag{2}$$

where $\chi^a_{bc}$ describes the second order susceptibility [46,47]. The electric field product can be decomposed into the real symmetric and imaginary antisymmetric parts,

$$\mathcal{E}_b(\omega) \mathcal{E}_c(-\omega) = \mathcal{E}_b(\omega) \mathcal{E}^*_c(\omega) = \Re(\mathcal{E}_b \mathcal{E}^*_c) + i\Im(\mathcal{E}_b \mathcal{E}^*_c), \tag{3}$$

$$\mathcal{E}_b(-\omega) \mathcal{E}_c(\omega) = \mathcal{E}^*_b(\omega) \mathcal{E}_c(\omega) = \Re(\mathcal{E}^*_b \mathcal{E}_c) - i\Im(\mathcal{E}^*_b \mathcal{E}_c). \tag{4}$$

The real part describes the LPL illumination, while the imaginary part corresponds to the electric field phase difference in CPL. According to the Kubo perturbation theory, one can evaluate the SAM variation under LPL according to (Supplementary Information Text I)

$$\delta m_S^{a,\text{LPL}} = \left[\eta_{\text{nor}}^{abc}(\omega) + \tau \frac{d\eta_{\text{mag}}^{abc}(\omega)}{dt}\right] 2\Re[\mathcal{E}_b(\omega)\mathcal{E}_c(-\omega)]. \tag{5}$$

There are two terms contributing to the susceptibility. The first term $\eta_{\text{nor}}^{abc}(\omega)$ is independent on $\tau$, in the form of

$$\eta_{\text{nor}}^{abc}(\omega) = \frac{g_S V_{\text{u.c.}} \pi e^2}{2\hbar^2} \int \frac{d^3 k}{(2\pi)^3} \sum_{m,n} f_{nm} \Im\left(r_{mn}^b \varsigma_{nm}^{c,a} + r_{mn}^c \varsigma_{nm}^{b,a}\right) \delta(\omega_{nm} - \omega). \tag{6}$$

Here, $f_{nm}$ and $\hbar\omega_{nm}$ measure the difference of occupation number and eigenenergy between states $n$ and $m$, respectively. $V_{\text{u.c.}}$ is the unit cell volume, and $g_S = 2$ is the spin Landé g-factor. The interband position matrix is $r_{mn}^b = \frac{v_{mn}^b}{i\omega_{mn}}, (m \neq n)$. We define $\varsigma_{nm}^{b,a}$ as

$$\varsigma_{nm}^{b,a} = \frac{i}{\omega_{nm}}\left[\frac{S_{nm}^a(v_{nn}^b - v_{mm}^b)}{\omega_{nm}} + \sum_{l \neq n,m}\left(\frac{v_{nl}^b S_{lm}^a}{\omega_{lm}} - \frac{S_{nl}^a v_{lm}^b}{\omega_{nl}}\right)\right]. \tag{7}$$

The terms in the hard bracket can be considered as a generalized spin derivative with respect to momentum $\boldsymbol{k}$. Similar as the shift current generation [44,48-53], $\eta_{\text{nor}}^{abc}(\omega)$ describes an intrinsic interband transition between carriers at the valence and conduction bands, which is independent on time.

The second contribution in Eq. (5) $\left[\frac{d\eta_{\text{mag}}^{abc}(\omega)}{dt}\right]$ evolves with time and saturates at the carrier lifetime $\tau$, which can be evaluated via

$$\frac{d\eta_{\text{mag}}^{abc}}{dt}(\omega) = -\frac{g_S V_{\text{u.c.}} \pi e^2}{2\hbar^2} \int \frac{d^3 k}{(2\pi)^3} \sum_{m,n} f_{nm} \Delta S_{nm}^a \{r_{mn}^b, r_{nm}^c\} \delta(\omega_{mn} - \omega). \tag{8}$$

Here, $\Delta S_{nm}^a = S_{nn}^a - S_{mm}^a$ is the intrinsic SAM difference between the valence and conduction bands. It is scaled by quantum metric tensor $g_{mn}^{bc} = \{r_{mn}^b, r_{nm}^c\} = r_{mn}^b r_{nm}^c + r_{mn}^c r_{nm}^b$, which is the real symmetric component of quantum geometric structure of Bloch wavefunction [54]. Hence, $\frac{d\eta_{\text{mag}}^{abc}}{dt}(\omega)$ describes the "velocity" of spin variation (similar as injection current generation [48,50,55-57] in the photocurrent process). One sees that the $\eta_{\text{nor}}^{abc}$ originates from the off-diagonal term of $S_{nm}^a$, while the $\frac{d\eta_{\text{mag}}^{abc}}{dt}$ evaluates its diagonal components contribution.

Alternatively, if a CPL is applied, the SAM contributed magnetic moment variation also contains two sources

$$\delta m_S^{a,\text{CPL}} = \left[\xi_{\text{mag}}^{abc}(\omega) + \tau \frac{d\xi_{\text{nor}}^{abc}(\omega)}{dt}\right] 2\Im[\mathcal{E}_b(\omega)\mathcal{E}_c(-\omega)], \qquad (9)$$

with

$$\xi_{\text{mag}}^{abc}(\omega) = \frac{g_S V_{\text{u.c.}} \pi e^2}{2\hbar^2} \int \frac{d^3 \mathbf{k}}{(2\pi)^3} \sum_{m,n} f_{nm} \Re\left(r_{mn}^b \varsigma_{nm}^{c,a} - r_{mn}^c \varsigma_{nm}^{b,a}\right)\delta(\omega_{nm} - \omega) \qquad (10)$$

and

$$\frac{d\xi_{\text{nor}}^{abc}}{dt}(\omega) = -\frac{ig_S V_{\text{u.c.}} \pi e^2}{2\hbar^2} \int \frac{d^3 \mathbf{k}}{(2\pi)^3} \sum_{m,n} f_{nm} \Delta S_{nm}^a [r_{mn}^b, r_{nm}^c]\delta(\omega_{mn} - \omega). \qquad (11)$$

Here, the commutator $[r_{mn}^b, r_{nm}^c] = r_{mn}^b r_{nm}^c - r_{mn}^c r_{nm}^b$ is proportional to the interband Berry curvature, serving as the imaginary antisymmetric part of quantum geometric nature of Bloch wavefunction. It is evident that this photoinduced SAM variation [Eqs. (5) and (9)] can occur under photon excitation between the valence and conduction bands in intrinsic semiconductors without a finite Fermi surface.

### B. Symmetry constraints on different susceptibilities

Next, we briefly discuss the symmetry arguments of these susceptibilities. Since $\eta_{\text{nor}}^{abc}$, $\frac{d\eta_{\text{mag}}^{abc}}{dt}$, $\xi_{\text{mag}}^{abc}$, and $\frac{d\xi_{\text{nor}}^{abc}}{dt}$ all compose one SAM (transforming as axial vectors) and two position (transforming as polar vectors) operators, it is obvious that they are all invariant under $\mathcal{P}$, namely, parity-even. Hence, it can be used in centrosymmetric systems. Note that this is significantly different from another second order optical effect, bulk photovoltaic effect, which requires $\mathcal{P}$-broken. As for $\mathcal{T}$, one can show that both $\frac{d\eta_{\text{mag}}^{abc}}{dt}$ and $\xi_{\text{mag}}^{abc}$ are zero for nonmagnetic materials and only arises in $\mathcal{T}$-broken systems, while $\eta_{\text{nor}}^{abc}$ and $\frac{d\xi_{\text{nor}}^{abc}}{dt}$ are proven to emerge regardless of $\mathcal{T}$ (see Supplementary Information Text II). We therefore add a subscript "mag" to the former two susceptibilities to reflect their $\mathcal{T}$-broken nature.

## C. Circularly polarized light effect on monolayer NiCl$_2$

Having elaborated the microscopic mechanism of photoinduced SAM variation, we now take a prototypical 2D ferromagnetic material, monolayer NiCl$_2$, as an exemplary platform and perform first-principles density functional theory [58,59] calculations to illustrate the process. Note that the family of monolayered nickel dihalides were synthesized recently and have drawn extensive interests due to their unique magnetic feature [36,60]. Each Ni$^{2+}$ ion carries 2 $\mu_B$ magnetic moments from SAM. The Bloch states do not exhibit net intrinsic OAM under inversion symmetry, but we will show later on that OAM also makes significant contributions to magnetization reorientation, equally important as that from SAM. The monolayer NiCl$_2$ possesses a strong ferromagnetic configuration, stabilized by biquadratic interactions [57], and prefers an in-plane orientation [36]. Interestingly, this 2D ferromagnet has also been revealed to exhibit another compelling optomagnetic feature known as half-excitonic insulator phase [61], which can mediate a new type of Bose-Einstein condensation.

The monolayer NiCl$_2$ exhibits a sandwiched atomic structure, with two Cl atomic layers on the both sides of the Ni atomic layer. The three triangular atomic layers form a 1$T$ phase in the trigonal lattice [Fig. 1(a), layer group of $P\bar{3}m1$], so that $\mathcal{P}$ preserves under collinear ferromagnetic configurations of any orientation. The spin-orbit coupling (SOC) effect breaks spin rotational symmetry, and the in-plane spin orientation is energetically lower than the out-of-plane pattern by ~0.01 meV per unit cell. Without loss of generality, we calculate the electronic band dispersion when the intrinsic magnetization on Ni (denoted by $\boldsymbol{m}_{\text{Ni}}$) along $x$, as plotted in Fig. 1(b). It shows a semiconducting feature with the calculated bandgap of 2.43 eV.

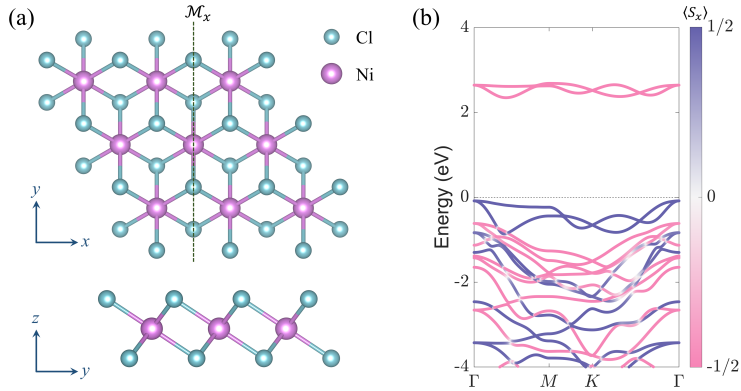

Figure 1. Geometric and electronic structure of monolayer NiCl$_2$. (a) Atomic structure from top and side views, with the $\mathcal{M}_x$ mirror reflection plane indicated. (b) Electronic band dispersion along

the high symmetric **k**-path calculated with the intrinsic $\boldsymbol{m}_{\text{Ni}}$ aligned along *x*, with the band-resolved $\langle S_x \rangle$ color-mapped.

We then calculate the photoinduced SAM variation susceptibility coefficients. We focus on CPL irradiation, and the LPL results are shown in Supplementary Information. As the NiCl$_2$ is $\mathcal{T}$-broken, both $\xi_{\text{mag}}(\omega)$ and $\frac{d\xi_{\text{nor}}(\omega)}{dt}$ should be present. Nonetheless, the specific magnetic point group imposes further constraints on their symmetry arguments. For example, when $\boldsymbol{m}_{\text{Ni}}$ is along *x* (breaking the $\mathcal{C}_{3z}$ rotation), the system belongs to the magnetic point group of $2/m$, which is invariant under $\mathcal{C}_{2x}$, $\mathcal{M}_x$, and $\mathcal{P}$ operations. One can apply a simple group theory analysis to determine the existence of susceptibility components. The CPL electric field components transform as $\Gamma_{\text{CPL}} = A_u \otimes B_u = B_g$. The spin-*x* (*y* and *z*) component follows $A_g$ ($B_g$). Hence, there will be no responses to change the *x* component, as $B_g \otimes A_g = B_g$ is symmetrically forbidden. The other two components are allowed, as the $B_g \otimes B_g = A_g$ gives the identical irreducible representation. One can clearly observe the quantitative results in Figs. 2(a) and 2(b) fully consistent with these symmetry analyses. Note that if the light handedness is flipped, the response functions reverse their sign, in accordance with Eqs. (10) and (11). Generally speaking, the $\frac{d\xi_{\text{nor}}(\omega)}{dt}$ plays a dominant role in dictating the spin polarization as compared with $\xi_{\text{mag}}(\omega)$, if a conservative carrier lifetime of ~0.1 ps is assumed. It can be seen that the z-component $\frac{d\xi_{\text{nor}}(\omega)}{dt}$ is in general larger than that along *y*. When the field strength reaches 0.1 V/nm, it generates ~$10^{-4}$ $\mu_B$ magnetic moment. This clearly infers that light could switch the SAM direction along the out-of-plane.

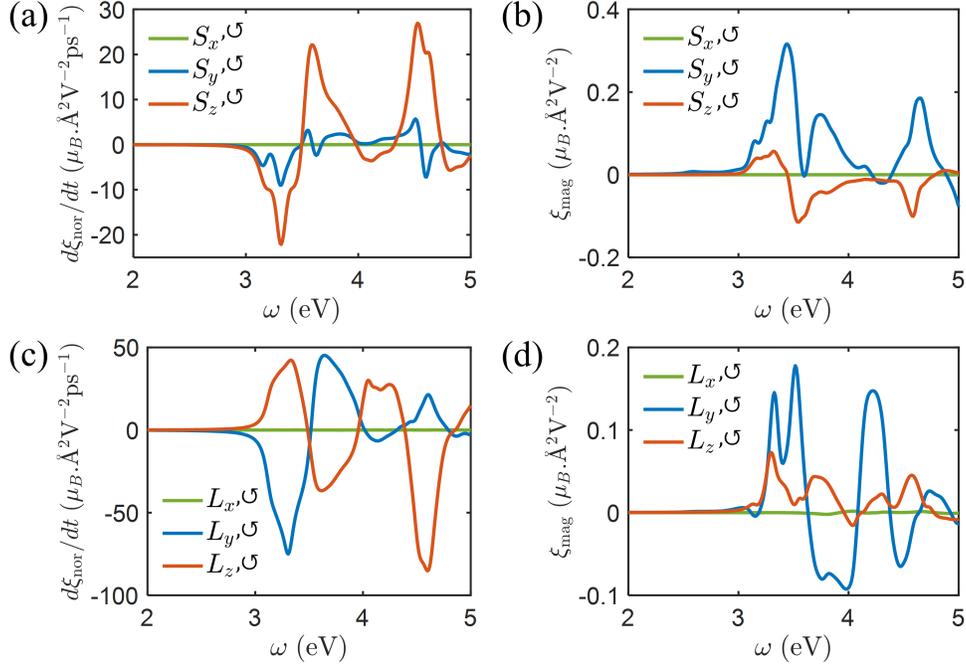

Figure 2. Left-handed CPL induced magnetization variation when $\boldsymbol{m}_{\text{Ni}}$ is along $+x$, namely, $\boldsymbol{m}_{\text{Ni}} = (2,0,0)\mu_B$. (a) and (b) plot SAM variations $\frac{d\xi^{S,\circlearrowleft}_{\text{nor}}(\omega)}{dt}$ and $\xi^{S,\circlearrowleft}_{\text{mag}}(\omega)$, as functions of incident photon energy, respectively. (c) and (d) are their OAM counterparts, $\frac{d\xi^{L,\circlearrowleft}_{\text{nor}}(\omega)}{dt}$ and $\xi^{L,\circlearrowleft}_{\text{mag}}(\omega)$, respectively.

Other than SAM, OAM also contributes to the magnetic moments. The OAM has been overlooked for a long time as its intrinsic value generally quenches in strong and symmetric crystals, such as in the current situation. Recently, the OAM based electronics (i.e., orbitronics) receives its renaissance [62-69]. The OAM conceives much higher information transport efficiency, longer distance, larger magnetic moments, and faster control kinetics, while its material platform does not require heavy elements with strong SOC effect. For example, it has been shown that the orbital Hall effect [70,71], an OAM version of spin Hall effect, could be used for generating and accumulating large magnetic moments for information process and developing terahertz light sources. The OAM contributions for magnetoelectric coupling (orbital Rashba-Edelstein effect) [46,72-74] and photocurrent (bulk orbital photovoltaic effect) [65] may overpass the SAM contributions in nonmagnetic and antiferromagnetic systems. In this regard, we compute the OAM counterpart by replacing the spin operators $\boldsymbol{S}$ by $\boldsymbol{L}$ (Supplementary Information Text III). The results are shown in Figs. 2(c) and 2(d). Note that both SAM and OAM transform as pseudovectors, therefore their symmetry constraints are the same. Hence, both $\frac{d\xi_{\text{nor}}^{L_x,\circlearrowleft}(\omega)}{dt}$ and $\xi_{\text{mag}}^{L_x,\circlearrowleft}(\omega)$ diminish under $\mathcal{M}_x$. For the symmetry-allowed terms, it is clear that the OAM and SAM contributions to magnetic moment susceptibility are on the same order of magnitude and neither can be omitted.

When the intrinsic magnetization is along another direction, the magnetic group varies to $\bar{3}m' = S_6 + (D_{3d} - S_6)\mathcal{T}$. This would change the symmetry arguments for their existence (Supplementary Information Text II). In Figure 3 we plot the responses when $\boldsymbol{m}_{\text{Ni}}$ is along out-of-plane $z$ direction. Similarly, the dominant factors are in Figs. 3(a) and 3(c), but one can see that the in-plane components of both $\frac{d\xi_{\text{nor}}^{\boldsymbol{S},\circlearrowleft}(\omega)}{dt}$ and $\frac{d\xi_{\text{nor}}^{\boldsymbol{S},\circlearrowleft}(\omega)}{dt}$ vanish, according to symmetry constraints. This suggests that when the magnetic moment is switched along out-of-plane, it would be pinned to this direction without further tilts.

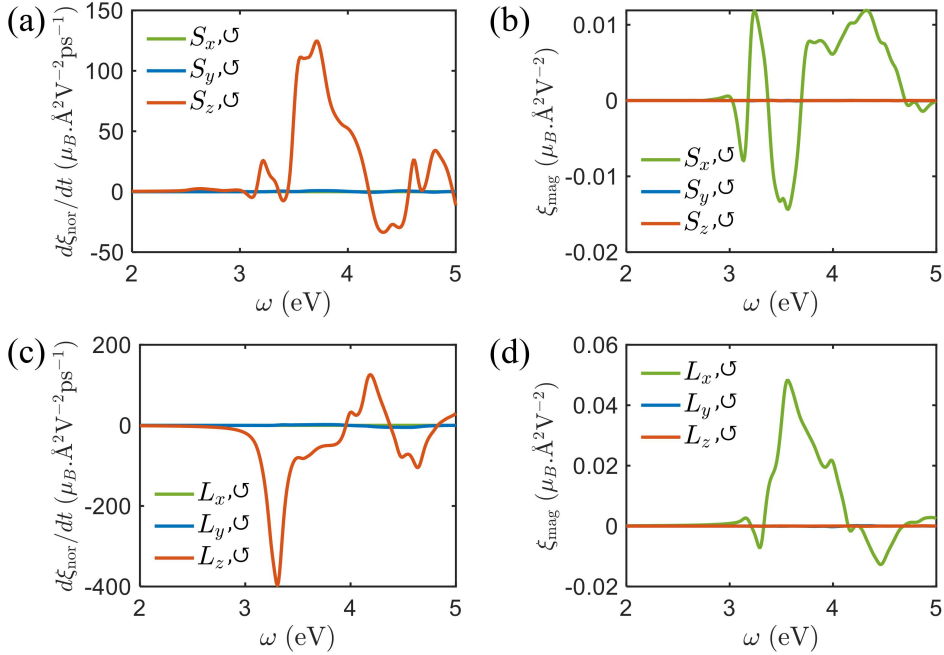

Figure 3. Left-handed CPL induced magnetization variation when $\boldsymbol{m}_{\text{Ni}}$ is along +$z$. (a) and (b) plot SAM variations $\frac{d\xi_{\text{nor}}^{\boldsymbol{S},\circlearrowleft}(\omega)}{dt}$ and $\xi_{\text{mag}}^{\boldsymbol{S},\circlearrowleft}(\omega)$, as functions of incident photon energy, respectively. (c) and (d) are their OAM counterparts, $\frac{d\xi_{\text{nor}}^{\boldsymbol{L},\circlearrowleft}(\omega)}{dt}$ and $\xi_{\text{mag}}^{\boldsymbol{L},\circlearrowleft}(\omega)$, respectively.

We now rotate $\boldsymbol{m}_{\text{Ni}}$ and plot the results in Fig. 4. Here, we sum the SAM and OAM contributions together. It can be seen that when $\boldsymbol{m}_{\text{Ni}}$ is lying in the horizontal $xy$ plane, the magnetization torque along $z$ $[\frac{d\xi_{\text{nor}}^{L_z+S_z,\circlearrowleft}(\omega)}{dt}]$ almost keeps a constant [Fig. 4(a)], while the torques along $x$ and $y$ mainly depends on $\sin 2\varphi$ and $\cos 2\varphi$, respectively. Once $\bm{m}_{\text{Ni}}$ is tilted off the $xy$ plane, the $z$ component torque is negative and becomes dominant over other components [Fig. 4(c)], which helps to vertically align the $\bm{m}_{\text{Ni}}$. These $\bm{m}_{\text{Ni}}$-dependent torques vividly suggest that we can regulate the in-plane magnetization to align with the out-of-plane direction via CPL illumination, which is highly desirable for the magnetic information storage in modern spintronic applications.

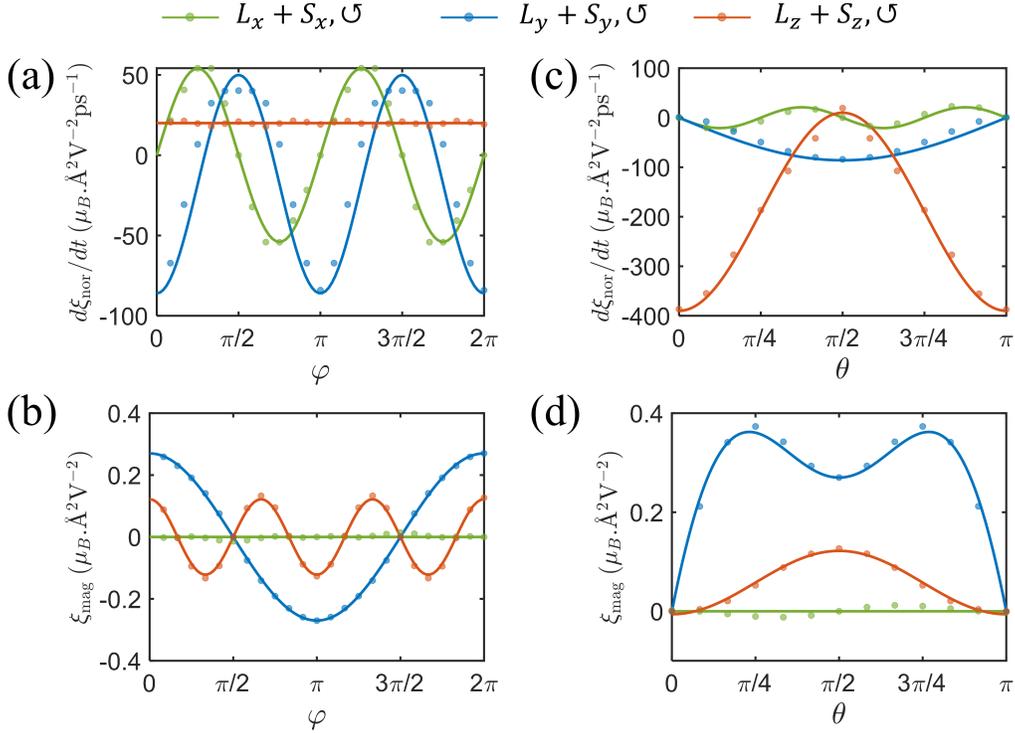

Figure 4. Direction dependence of magnetization on Ni. (a) and (b) are $\frac{d\xi_{\text{nor}}^{L+S,\circlearrowleft}(\omega)}{dt}$ and $\xi_{\text{mag}}^{L+S,\circlearrowleft}(\omega)$ when the $\bm{m}_{\text{Ni}}$ is in the $xy$ plane (polar angle $\theta = \frac{\pi}{2}$), as a function of the azimuthal angle $\varphi$, respectively. (c) and (d) plot the $\frac{d\xi_{\text{nor}}^{L+S,\circlearrowleft}(\omega)}{dt}$ and $\xi_{\text{mag}}^{L+S,\circlearrowleft}(\omega)$ when $\bm{m}_{\text{Ni}}$ is in the $xz$ plane ($\varphi = 0$), depending on $\theta$. We take an incident photon energy of $\hbar\omega = 3.3$ eV. The scatters are numerically results, and the curves are fitted by trigonometric functions. Note that the carrier relaxation time is typically on the order of 0.1 ps, hence $\frac{d\xi_{\text{nor}}^{L+S,\circlearrowleft}(\omega)}{dt}$ usually surpasses $\xi_{\text{mag}}^{L+S,\circlearrowleft}(\omega)$.

The $\mathcal{T}$ and $\mathcal{M}_x$ further pose additional symmetry constraints on the rotation of $\boldsymbol{m}_{\text{Ni}}$ with respect to $\varphi$ and $\theta$, namely, $\mathcal{T}:(\varphi,\theta) \to (\varphi+\pi, \pi-\theta)$ and $\mathcal{M}_x:(\varphi,\theta) \to (2\pi-\varphi, \pi-\theta)$. Note that $\frac{d\xi_{\text{nor}}}{dt}$ ($\xi_{\text{mag}}$) is allowed (forbidden) under $\mathcal{T}$, and would show different trigonometric function forms under spin rotation. For example, $\frac{d\xi_{\text{nor}}}{dt}$ at its lower order expansion is compatible with terms like $\sin\theta$, $\sin 2\varphi$, $\cos 2\theta$, $\cos 2\varphi$, and a constant $C$, but not $\sin\varphi$, $\cos\varphi$, $\cos\theta$ or $\sin 2\theta$, at its lower order expansion. The $\xi_{\text{mag}}$ takes the opposite manner. Similar analysis can be applied for $\mathcal{M}_x$, which flips $\frac{d\xi_{\text{nor}}^x}{dt}$ and $\xi_{\text{mag}}^x$, and leaves $\frac{d\xi_{\text{nor}}^{y/z}}{dt}$ and $\xi_{\text{mag}}^{y/z}$ invariant. We accordingly use polynomial functions to fit the angle dependence of $\frac{d\xi_{\text{nor}}}{dt}$ and $\xi_{\text{mag}}$ in the lower order, and the detailed results can be seen in Supplementary Information Text IV.

### D. Estimate of collective spin reorientation

We further perform a simplified magnetization dynamics simulation to illustrate the reorientation process of $\boldsymbol{m}_{\text{Ni}}$, according to the Landau–Lifshitz–Gilbert equation [39]

$$\frac{\partial \widetilde{\boldsymbol{m}}}{\partial t} = -\gamma \widetilde{\boldsymbol{m}} \times \boldsymbol{B}_{\text{eff}} + \alpha \widetilde{\boldsymbol{m}} \times \frac{\partial \widetilde{\boldsymbol{m}}}{\partial t} + \boldsymbol{T} \tag{12}$$

Here, $\widetilde{\boldsymbol{m}} = \frac{\boldsymbol{m}_{\text{Ni}}}{|\boldsymbol{m}_{\text{Ni}}|}$ is the unit vector of the intrinsic magnetic moment, $\gamma$ and $\alpha$ are gyromagnetic ratio and Gilbert damping constant, respectively, describing the precession and nutation motions. $\boldsymbol{B}_{\text{eff}}$ is the effective magnetic field that originates from both inherent magnetic exchange couplings ($\boldsymbol{B}_{\text{eff}} = -\frac{\partial \mathcal{H}}{\partial \boldsymbol{m}}$ with $\mathcal{H}$ the Heisenberg model Hamiltonian) and the $\tau$-independent photomagnetization effect, $\boldsymbol{B}_{\text{eff}} = J_{ex}\xi_{\text{mag}}E^2/\mu_B^2$ ($J_{ex}$ is exchange parameter and $E$ is field strength). Finally, noting that $\tau$ is much longer than the simulation time step (taken to be 0.65 fs here), the effect from magnetic variation rate over time $\frac{d\xi_{\text{nor}}}{dt}$ is taken into account via an effective time-dependent magnetic torque $\boldsymbol{T}$ ($= \frac{d\xi_{\text{nor}}}{dt}E^2$), depending on time $t$. Limited by the optical aperture and its long wavelength nature, it is reasonable to assume a homogenous light illuminates on the sample. In this regard, we only consider coherent $\boldsymbol{m}_{\text{Ni}}$ reorientation motions and ignore the magnon generation.

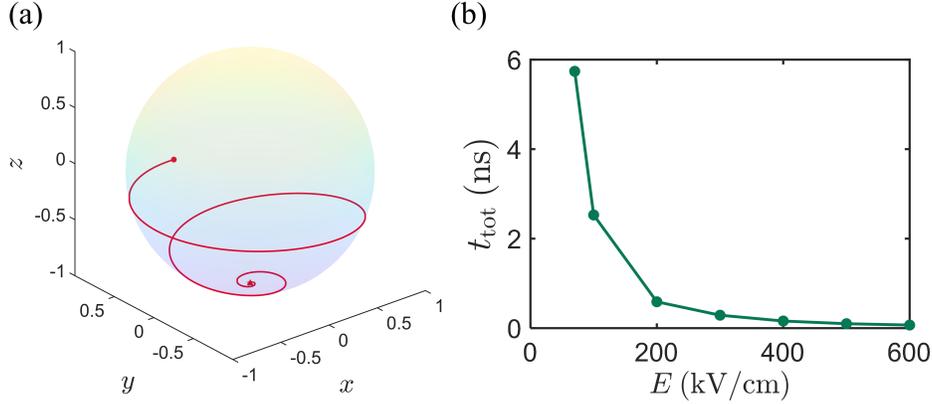

Figure 5. The magnetization dynamics under CPL. (a) The initial magnetization is along *y* and it swirls to −*z* during simulation and pins at $\theta = \pi$. The light electric field strength is 70 kV/cm. (b) Variation of total time needed to vertically align $\boldsymbol{m}_{\text{Ni}}$ as a function of incident electric field strength. The damping factor is set to a conservative value $\alpha = 0.05$. The incident photon energy is 3.3 eV, or a wavelength of 375 nm.

As sketched by the trajectory in Fig. 5(a), our magnetic dynamics simulations show that $\boldsymbol{m}_{\text{Ni}}$ will ultimately point along −*z* direction, starting from an initial position around the equator. We use the light electric field magnitude of 70 kV/cm (or a light intensity of 6.5 MW/cm²), and the reorientation can be accomplished within 5.7 ns. Note that if the light field is enhanced, the switching time can be significantly reduced [Fig. 5(b)]. If the light field is increased to 200 kV/cm, it takes only 0.6 ns to reorient the magnetization. Such timescale renders CPL illumination an efficient nonlinear optical approach to achieve ultrafast magnetization control and manipulation, with only modest light intensity required. Note that in the current case, we do not include excitonic interactions, vertex corrections, and disorder effects, which do not alter the main conclusions and the symmetry arguments.

### E. Linearly polarized light effect to switch spin

Apart from CPL, our theory suggests that the LPL irradiation also induces magnetic torque $\boldsymbol{T}$ and effective magnetic field $\boldsymbol{B}_{\text{eff}}$ in magnetic systems, both of which can be used to adjust the magnetization orientation as well. In this circumstance, the dominant term will be $\frac{d\eta_{\text{mag}}}{dt}$. If the magnetization is strictly in the *xy* plane, the out-of-plane torque would be vanishingly small.

Nonetheless, once it is slightly tilted off-plane (such as by thermal excitation), a pronounced $z$ component arises in the torque and overwhelms the other two components. Then one could also expect an out-of-plane magnetization ultimately (see Supplementary Information Figures S1–S3 and Figure 6). Note that this is a bit counterintuitive. Unlike CPL that is usually treated as an effective magnetic field along its propagating direction, LPL does not break $\mathcal{T}$. We have revealed that this effective torque microscopically arises from the large quantum metric, which scales with $\frac{d\eta_{\text{mag}}}{dt}$. With these insights, we perform a dynamics simulation, and show that such an out-of-plane switching can also be accomplished in an ultrafast timescale.

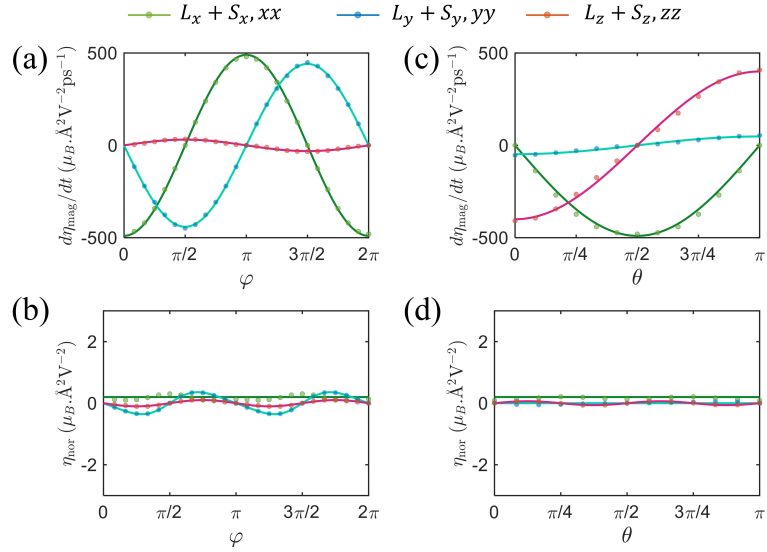

Figure 6. Direction dependence of spin on Ni under $x$-polarized LPL. (a) and (b) are $\frac{d\xi_{\text{nor}}^{L+S,xx}(\omega)}{dt}$ and $\xi_{\text{mag}}^{L+S,xx}(\omega)$ when the Ni spin is in the $x$–$y$ plane (polar angle $\theta = \frac{\pi}{2}$), as a function of azimuthal angle $\varphi$, respectively. (c) and (d) plot the $\frac{d\eta_{\text{mag}}^{L+S,xx}(\omega)}{dt}$ and $\eta_{\text{nor}}^{L+S,xx}(\omega)$ when spin is in the $x$–$z$ plane ($\varphi = 0$), depending on $\theta$. We take an incident photon energy of $\hbar\omega = 3.5$ eV. The scatters are calculated values, and the curves are their trigonometric fitting results.

**Conclusion**

In conclusion, we have developed a general band formulism to describe the nonlinear optical effect on the SAM and OAM variations, which could yield a change of intrinsic magnetization and its switching velocity, or generate magnetism in nonmagnetic materials. As a parity-even process, such light-assisted magnetization switching can take place on centrosymmetric semiconductors.

Moreover, the OAM effect in this context does not rely on strong spin-orbit coupling, hence can be detected even in materials only composed of light-weight elements. By applying the theory to a prototypical monolayer ferromagnet $NiCl_2$, we show that light irradiation could collectively lift the intrinsic magnetization from its easy-magnetization plane to the out-of-plane direction within a few nanoseconds. This light-induced magnetization switching has not constraint on the magnetic ground state and can also be achieved in magnetic and nonmagnetic systems. Our work opens a distinctive route to effectively regulating magnetization of condensed matter via noncontacting optical illumination, which should prove instrumental for ultrafast control and quantum manipulation.

**Methods**

We use first-principles density functional theory (DFT) calculations to perform numerical evaluations of the geometric, electronic, and magnetic properties of monolayer $NiCl_2$, which are implemented in the Vienna *ab initio* simulation package [75]. The exchange-correlation interactions are treated using the generalized gradient approximation (GGA) method in the solid state Perdew–Burke–Enzerhof (PBE) form [76]. Projector augmented-wave [77] and planewave basis set are used to describe the core and valence electrons, respectively. In order to treat the strong correlation of Ni *d* orbitals, we add Hubbard *U* corrections with its effective value setting to be 4 eV. We have tested other values which do not qualitatively change the results. The kinetic energy cutoff is set to be 400 eV, and convergence criteria for the self-consistent field total energy and force component are set to be $1\times10^{-7}$ eV and $1\times10^{-3}$ eV/Å, respectively. A vacuum space in the z direction of 15 Å is added to eliminate the layer image interactions, and the Monkhorst-Pack *k*-meshes [78] of (18×18×1) grids are used to perform integration in the first Brillouin zone. Self-consistent SOC interactions are always included in the calculations.

**Acknowledgments.** The work is supported by National Natural Science Foundation of China (NSFC) under Grants No. 12374065. The calculations are performed on the HPC platform of Xi'an Jiaotong University, the National Supercomputing Center in Xi'an, and the Hefei Advanced Computing Center.

**References:**

[1]  S.-i. Ohkoshi, Y. Tsunobuchi, T. Matsuda, K. Hashimoto, A. Namai, F. Hakoe, and H. Tokoro, Synthesis of a metal oxide with a room-temperature photoreversible phase transition, Nat. Chem. **2**, 539 (2010).

[2]  Y. Okimoto, X. Peng, M. Tamura *et al.*, Ultrasonic Propagation of a Metallic Domain in $Pr_{0.5}Ca_{0.5}CoO_3$ Undergoing a Photoinduced Insulator-Metal Transition, Phys. Rev. Lett. **103**, 027402 (2009).

[3]  Y. Matsubara, S. Ogihara, J. Itatani *et al.*, Coherent dynamics of photoinduced phase formation in a strongly correlated organic crystal, Phys. Rev. B **89**, 161102 (2014).

[4]  J. Zhou, H. Xu, Y. Shi, and J. Li, Terahertz Driven Reversible Topological Phase Transition of Monolayer Transition Metal Dichalcogenides, Adv. Sci. **8**, 2003832 (2021).

[5]  V. D. Mihailetchi, J. Wildeman, and P. W. M. Blom, Space-Charge Limited Photocurrent, Phys. Rev. Lett. **94**, 126602 (2005).

[6]  Y. Li, J. Fu, X. Mao, C. Chen, H. Liu, M. Gong, and H. Zeng, Enhanced bulk photovoltaic effect in two-dimensional ferroelectric $CuInP_2S_6$, Nat. Commun. **12**, 5896 (2021).

[7]  W. Ji, K. Yao, and Y. C. Liang, Bulk Photovoltaic Effect at Visible Wavelength in Epitaxial Ferroelectric $BiFeO_3$ Thin Films, Adv. Mater. **22**, 1763 (2010).

[8]  W. T. H. Koch, R. Munser, W. Ruppel, and P. Würfel, Bulk photovoltaic effect in $BaTiO_3$, Solid State Commun. **17**, 847 (1975).

[9]  F. Siegrist, J. A. Gessner, M. Ossiander *et al.*, Light-wave dynamic control of magnetism, Nature **571**, 240 (2019).

[10] D. Afanasiev, J. R. Hortensius, B. A. Ivanov, A. Sasani, E. Bousquet, Y. M. Blanter, R. V. Mikhaylovskiy, A. V. Kimel, and A. D. Caviglia, Ultrafast control of magnetic interactions via light-driven phonons, Nat. Mater. **20**, 607 (2021).

[11] M. Burresi, D. van Oosten, T. Kampfrath, H. Schoenmaker, R. Heideman, A. Leinse, and L. Kuipers, Probing the Magnetic Field of Light at Optical Frequencies, Science **326**, 550 (2009).

[12] A. D. Kent and D. C. Worledge, A new spin on magnetic memories, Nat. Nanotechnol. **10**, 187 (2015).

[13] T. Kawahara, K. Ito, R. Takemura, and H. Ohno, Spin-transfer torque RAM technology: Review and prospect, Microelectron. Reliab. **52**, 613 (2012).

[14] B. Dieny, I. L. Prejbeanu, K. Garello *et al.*, Opportunities and challenges for spintronics in the microelectronics industry, Nat. Electron. **3**, 446 (2020).

[15] P. Pirro, V. I. Vasyuchka, A. A. Serga, and B. Hillebrands, Advances in coherent magnonics, Nat. Rev. Mater. **6**, 1114 (2021).

[16] Y.-C. Lau, D. Betto, K. Rode, J. M. D. Coey, and P. Stamenov, Spin–orbit torque switching without an external field using interlayer exchange coupling, Nat. Nanotechnol. **11**, 758 (2016).

[17] A. El-Ghazaly, J. Gorchon, R. B. Wilson, A. Pattabi, and J. Bokor, Progress towards ultrafast spintronics applications, J. Magn. Magn. Mater. **502**, 166478 (2020).

[18] S. Kumar and S. Kumar, Ultrafast light-induced THz switching in exchange-biased Fe/Pt spintronic heterostructure, Appl. Phys. Lett. **120**, 202403 (2022).

[19] C. Zheng, K. Zhu, S. C. d. Freitas *et al.*, Magnetoresistive Sensor Development Roadmap (Non-Recording Applications), IEEE Trans. Magn. **55**, 1 (2019).

[20] P. P. Freitas, R. Ferreira, and S. Cardoso, Spintronic Sensors, Proc. IEEE **104**, 1894 (2016).

[21] A. V. Silva, D. C. Leitao, J. Valadeiro, J. Amaral, P. P. Freitas, and S. Cardoso, Linearization strategies for high sensitivity magnetoresistive sensors, Eur. Phys. J. Appl. Phys. **72**, 10601 (2015).

[22] R. C. Chaves, P. P. Freitas, B. Ocker, and W. D. Maass, Low frequency picotesla field detection using hybrid MgO based tunnel sensors, Appl. Phys. Lett. **91**, 102504 (2007).

[23]   O. Ben Dor, S. Yochelis, A. Radko *et al.*, Magnetization switching in ferromagnets by adsorbed chiral molecules without current or external magnetic field, Nat. Commun. **8**, 14567 (2017).

[24]   P. Wadley, B. Howells, J. Železný *et al.*, Electrical switching of an antiferromagnet, Science **351**, 587 (2016).

[25]   A. Hirohata, K. Yamada, Y. Nakatani, I.-L. Prejbeanu, B. Diény, P. Pirro, and B. Hillebrands, Review on spintronics: Principles and device applications, J. Magn. Magn. Mater. **509**, 166711 (2020).

[26]   V. M. Edelstein, Spin polarization of conduction electrons induced by electric current in two-dimensional asymmetric electron systems, Solid State Commun. **73**, 233 (1990).

[27]   Q. Xie, W. Lin, J. Liang *et al.*, Rashba–Edelstein Effect in the h-BN Van Der Waals Interface for Magnetization Switching, Adv. Mater. **34**, 2109449 (2022).

[28]   K. Ando and E. Saitoh, Observation of the inverse spin Hall effect in silicon, Nat. Commun. **3**, 629 (2012).

[29]   B. F. Miao, S. Y. Huang, D. Qu, and C. L. Chien, Inverse Spin Hall Effect in a Ferromagnetic Metal, Phys. Rev. Lett. **111**, 066602 (2013).

[30]   J. P. van der Ziel, P. S. Pershan, and L. D. Malmstrom, Optically-Induced Magnetization Resulting from the Inverse Faraday Effect, Phys. Rev. Lett. **15**, 190 (1965).

[31]   M. Battiato, G. Barbalinardo, and P. M. Oppeneer, Quantum theory of the inverse Faraday effect, Phys. Rev. B **89**, 014413 (2014).

[32]   A. Ben-Amar Baranga, R. Battesti, M. Fouché, C. Rizzo, and G. L. J. A. Rikken, Observation of the inverse Cotton-Mouton effect, Europhys. Lett. **94**, 44005 (2011).

[33]   D. Ejlli, On the CMB circular polarization: I. The Cotton–Mouton effect, Eur. Phys. J. C **79**, 231 (2019).

[34]   J. A. Armstrong, N. Bloembergen, J. Ducuing, and P. S. Pershan, Interactions between Light Waves in a Nonlinear Dielectric, Phys. Rev. **127**, 1918 (1962).

[35]   N. Bloembergen and P. S. Pershan, Light Waves at the Boundary of Nonlinear Media, Phys. Rev. **128**, 606 (1962).

[36]   M. A. McGuire, Crystal and Magnetic Structures in Layered, Transition Metal Dihalides and Trihalides, Crystals **7** (2017).

[37] D. I. Khomskii and S. V. Streltsov, Orbital Effects in Solids: Basics, Recent Progress, and Opportunities, Chem. Rev. **121**, 2992 (2021).

[38] Y.-G. Choi, D. Jo, K.-H. Ko *et al.*, Observation of the orbital Hall effect in a light metal Ti, Nature **619**, 52 (2023).

[39] T. L. Gilbert, A phenomenological theory of damping in ferromagnetic materials, IEEE Trans. Magn. **40**, 3443 (2004).

[40] H. Li, H. Gao, L. P. Zârbo *et al.*, Intraband and interband spin-orbit torques in noncentrosymmetric ferromagnets, Phys. Rev. B **91**, 134402 (2015).

[41] X. Li, H. Chen, and Q. Niu, Out-of-plane carrier spin in transition-metal dichalcogenides under electric current, Proc. Natl. Acad. Sci. U.S.A. **117**, 16749 (2020).

[42] A. G. Aronov and Y. B. Lyanda-Geller, Spin-orbit Berry phase in conducting rings, Phys. Rev. Lett. **70**, 343 (1993).

[43] E. M. Chudnovsky, Theory of Spin Hall Effect: Extension of the Drude Model, Phys. Rev. Lett. **99**, 206601 (2007).

[44] R. von Baltz and W. Kraut, Theory of the bulk photovoltaic effect in pure crystals, Phys. Rev. B **23**, 5590 (1981).

[45] B. I. Sturman, Ballistic and shift currents in the bulk photovoltaic effect theory, Phys. Usp. **63**, 407 (2020).

[46] Y. Sun, X. Mu, Q. Xue, and J. Zhou, Tailoring Photoinduced Nonequilibrium Magnetizations in $In_2Se_3$ Bilayers Adv. Opt. Mater. **10**, 2270057 (2022).

[47] J. Zhou, Photo-magnetization in two-dimensional sliding ferroelectrics, npj 2D Mater. Appl. **6**, 15 (2022).

[48] X. Mu, Q. Xue, Y. Sun, and J. Zhou, Magnetic proximity enabled bulk photovoltaic effects in van der Waals heterostructures, Phys. Rev. Res. **5**, 013001 (2023).

[49] Z. Qian, J. Zhou, H. Wang, and S. Liu, Shift current response in elemental two-dimensional ferroelectrics, npj Comput. Mater. **9**, 67 (2023).

[50] J. E. Sipe and A. I. Shkrebtii, Second-order optical response in semiconductors, Phys. Rev. B **61**, 5337 (2000).

[51] L. Z. Tan, F. Zheng, S. M. Young, F. Wang, S. Liu, and A. M. Rappe, Shift current bulk photovoltaic effect in polar materials—hybrid and oxide perovskites and beyond, npj Comput. Mater. **2**, 16026 (2016).

[52] D. E. Parker, T. Morimoto, J. Orenstein, and J. E. Moore, Diagrammatic approach to nonlinear optical response with application to Weyl semimetals, Phys. Rev. B **99**, 045121 (2019).

[53] R. Fei, L. Z. Tan, and A. M. Rappe, Shift-current bulk photovoltaic effect influenced by quasiparticle and exciton, Phys. Rev. B **101**, 045104 (2020).

[54] J. P. Provost and G. Vallee, Riemannian structure on manifolds of quantum states, Commun. Math. Phys. **76**, 289 (1980).

[55] H. Wang and X. Qian, Electrically and magnetically switchable nonlinear photocurrent in PT-symmetric magnetic topological quantum materials, npj Comput. Mater. **6**, 199 (2020).

[56] Y. Shi and J. Zhou, Coherence control of directional nonlinear photocurrent in spatially symmetric systems, Phys. Rev. B **104**, 155146 (2021).

[57] R. Fei, W. Song, L. Pusey-Nazzaro, and L. Yang, PT-Symmetry-Enabled Spin Circular Photogalvanic Effect in Antiferromagnetic Insulators, Phys. Rev. Lett. **127**, 207402 (2021).

[58] P. Hohenberg and W. Kohn, Inhomogeneous Electron Gas, Phys. Rev. **136**, B864 (1964).

[59] W. Kohn and L. J. Sham, Self-Consistent Equations Including Exchange and Correlation Effects, Phys. Rev. **140**, A1133 (1965).

[60] K. Riedl, D. Amoroso, S. Backes *et al.*, Microscopic origin of magnetism in monolayer 3d transition metal dihalides, Phys. Rev. B **106**, 035156 (2022).

[61] Z. Jiang, Y. Li, W. Duan, and S. Zhang, Half-Excitonic Insulator: A Single-Spin Bose-Einstein Condensate, Phys. Rev. Lett. **122**, 236402 (2019).

[62] M. J. Padgett, Orbital angular momentum 25 years on, Opt. Express **25**, 11265 (2017).

[63] B. A. Bernevig, T. L. Hughes, and S.-C. Zhang, Orbitronics: The Intrinsic Orbital Current in p-Doped Silicon, Phys. Rev. Lett. **95**, 066601 (2005).

[64] D. Go, D. Jo, C. Kim, and H.-W. Lee, Intrinsic Spin and Orbital Hall Effects from Orbital Texture, Phys. Rev. Lett. **121**, 086602 (2018).

[65] X. Mu, Y. Pan, and J. Zhou, Pure bulk orbital and spin photocurrent in two-dimensional ferroelectric materials, npj Comput. Mater. **7**, 61 (2021).

[66] D. Go, D. Jo, H.-W. Lee, M. Kläui, and Y. Mokrousov, Orbitronics: Orbital currents in solids, Europhys. Lett. **135**, 37001 (2021).

[67] M. Costa, B. Focassio, L. M. Canonico, T. P. Cysne, G. R. Schleder, R. B. Muniz, A. Fazzio, and T. G. Rappoport, Connecting Higher-Order Topology with the Orbital Hall Effect in Monolayers of Transition Metal Dichalcogenides, Phys. Rev. Lett. **130**, 116204 (2023).

[68] L. Salemi and P. M. Oppeneer, Theory of magnetic spin and orbital Hall and Nernst effects in bulk ferromagnets, Phys. Rev. B **106**, 024410 (2022).

[69] D. Lee, D. Go, H.-J. Park *et al.*, Orbital torque in magnetic bilayers, Nat. Commun. **12**, 6710 (2021).

[70] H. Kontani, T. Tanaka, D. S. Hirashima, K. Yamada, and J. Inoue, Giant Orbital Hall Effect in Transition Metals: Origin of Large Spin and Anomalous Hall Effects, Phys. Rev. Lett. **102**, 016601 (2009).

[71] D. Jo, D. Go, and H.-W. Lee, Gigantic intrinsic orbital Hall effects in weakly spin-orbit coupled metals, Phys. Rev. B **98**, 214405 (2018).

[72] H. Xu, J. Zhou, H. Wang, and J. Li, Light-induced static magnetization: Nonlinear Edelstein effect, Phys. Rev. B **103**, 205417 (2021).

[73] T. Yoda, T. Yokoyama, and S. Murakami, Orbital Edelstein Effect as a Condensed-Matter Analog of Solenoids, Nano Lett. **18**, 916 (2018).

[74] L. Salemi, M. Berritta, A. K. Nandy, and P. M. Oppeneer, Orbitally dominated Rashba-Edelstein effect in noncentrosymmetric antiferromagnets, Nat. Commun. **10**, 5381 (2019).

[75] G. Kresse and J. Furthmüller, Efficient iterative schemes for ab initio total-energy calculations using a plane-wave basis set, Phys. Rev. B **54**, 11169 (1996).

[76] J. P. Perdew, A. Ruzsinszky, G. I. Csonka, O. A. Vydrov, G. E. Scuseria, L. A. Constantin, X. Zhou, and K. Burke, Restoring the Density-Gradient Expansion for Exchange in Solids and Surfaces, Phys. Rev. Lett. **100**, 136406 (2008).

[77] P. E. Blöchl, Projector augmented-wave method, Phys. Rev. B **50**, 17953 (1994).

[78] H. J. Monkhorst and J. D. Pack, Special points for Brillouin-zone integrations, Phys. Rev. B **13**, 5188 (1976).